\documentclass[
twocolumn,
superscriptaddress,amsmath,amssymb
]{revtex4-1}
\usepackage{amsfonts}
\usepackage{bm}
\usepackage{color}

\usepackage{graphicx}
\usepackage{epstopdf}

\begin{document}
\title{Vibrational and magnetic properties of crystalline CuTe$_2$O$_5$}
\author{Y.V. Lysogorskiy}
\affiliation{Institute of Physics, Kazan Federal University, Kremlevskaya St.~16a, 420008 Kazan, Russia}
\author{R.M. Eremina}
\affiliation{Institute of Physics, Kazan Federal University, Kremlevskaya St.~16a, 420008 Kazan, Russia}
\affiliation{E. K. Zavoisky Physical-Technical Institute of RAS, 420029 Kazan, Russia}
\author{T.P. Gavrilova}
\affiliation{Institute of Physics, Kazan Federal University, Kremlevskaya St.~16a, 420008 Kazan, Russia}
\affiliation{E. K. Zavoisky Physical-Technical Institute of RAS, 420029 Kazan, Russia}
\author{O.V. Nedopekin}
\affiliation{Institute of Physics, Kazan Federal University, Kremlevskaya St.~16a, 420008 Kazan, Russia}
\author{D.A. Tayurskii}
\affiliation{Institute of Physics, Kazan Federal University, Kremlevskaya St.~16a, 420008 Kazan, Russia}
\affiliation{Centre for Quantum Technologies, Kazan Federal University, Kremlevskaya St.~16a, 420008 Kazan, Russia}
\date{\today}

\begin{abstract}
In the present work we have performed an \emph{ab initio} calculation of vibrational properties of CuTe$_2$O$_5$ by means of density functional theory method. One has compared calculated values with known experimental data on Raman and infrared spectroscopy in order to verify the obtained results. Lattice contribution to the heat capacity, obtained from the \emph{ab initio} simulations was added to magnetic contribution calculated from the simple spin hamiltonian model in order to obtain total heat capacity. Obtained result are in good agreement to the experimental data. Thus, the DFT methods could complement the experimental and theoretical studying of low-dimensional magnetic systems such as CuTe$_2$O$_5$.
\end{abstract}

\maketitle
\section{Introduction}

The low-dimensional magnet CuTe$_2$O$_5$ has attracted much attention because of the dispute concerning the magnetic structure of this compound.  One-dimensional magnetic structure was suggested based on ESR measurements~\cite{Eremina2008Anisotropic,Gavrilova2008Anisotropic}, while the two dimensional model was obtained from first-principles electronic structure calculations~\cite{das2008proposed,ushakov2009electronic} and heat capacity measurements~\cite{Eremina2011Magnetization}.  

In the present work we have performed an \emph{ab initio} calculations of lattice vibrational properties of crystalline CuTe$_2$O$_5$, and made a comparison with such experimental data, as Raman and infrared (IR) spectra in order to verify obtained results.  On the next step, the magnetic contribution to the heat capacity, calculated from alternating spin chain model (present work) and 2D-coupled dimer model~\cite{das2008proposed} were added to lattice contribution in order to obtain total heat capacity and compare it to the known experimental data~\cite{Eremina2011Magnetization}.

\section{Computational details}
 \begin{figure}
 \includegraphics[width=0.95\linewidth]{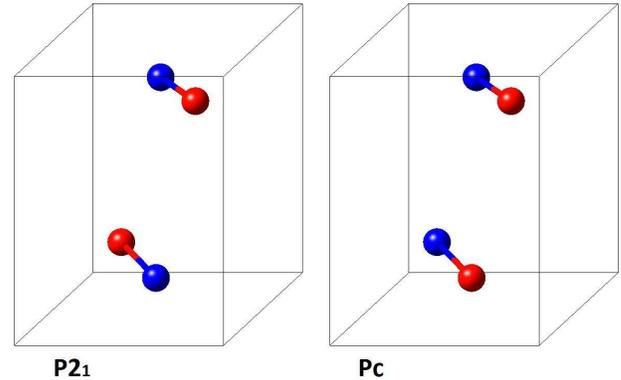}
 \caption{\label{fig:P21-Pc} The scheme of magnetic order of Cu-Cu structural dimers for two different symmetry subgroups (P2$_1$ and Pc) of space group P2$_1$/c. Red and blue spheres corresponds to the copper ions with spin up and spin down correspondingly, the magnetic moments value is $0.75$\,$\mu_\textrm{B}$.}
 \end{figure}
 
 \begin{figure*}
 \includegraphics[width=0.95\linewidth]{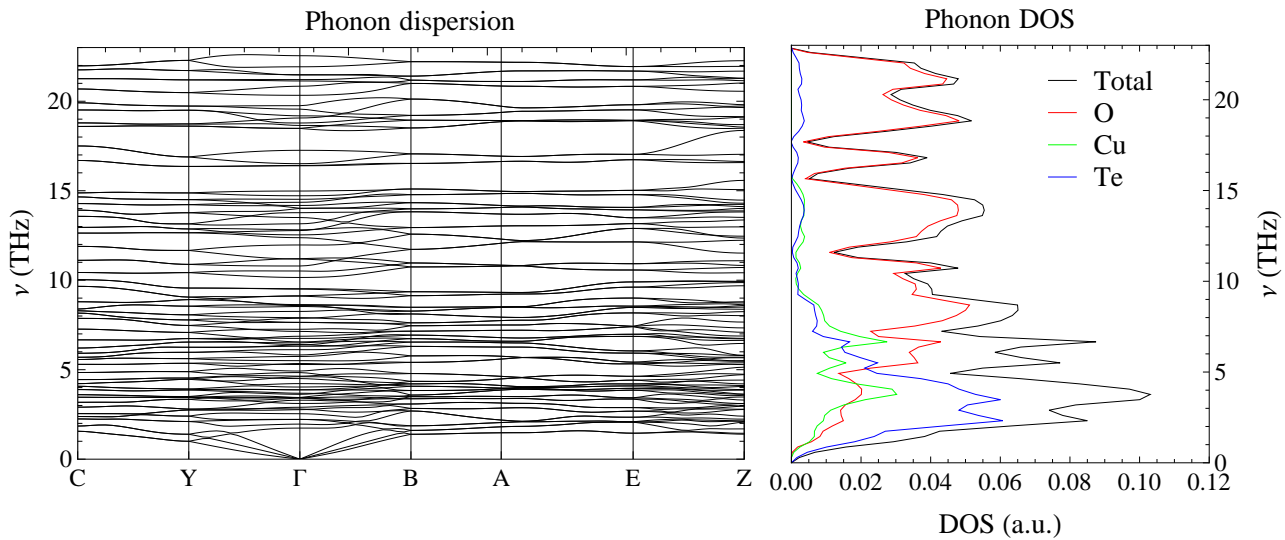}
 \caption{\label{fig:PhononDOS} Phonon dispersion and density of states, projected onto Cu, Te and O atoms for crystalline CuTe$_2$O$_5$.}
 \end{figure*}
The spin-polarized calculations were performed within the framework of density functional theory (DFT) with gradient-corrected exchange and correlation energy functional proposed by Perdew-Burke-Ernzerhof (PBE)~\cite{perdew1996generalized} and projector augmented-wave method as implemented in the code VASP~\cite{kresse1996efficient,kresse1999ultrasoft} (a part of the MedeA software package~\cite{MedeA}). The O($2s^22p^4$), Cu($3d^{10}4s^1$) and Te($5s^25p^4$) electrons were treated explicitly, whereas the rest were considered through the pseudopotentials. The maximum energy for plane wave basis set was selected to be equal 400 eV.
The electronic strong correlations were treated with the simplified (rotationally invariant) GGA+U approach, introduced by Dudarev \textit{et al.}~\cite{Dudarev}. The Coulomb on-site repulsion parameter U for Cu $d$-electrons was set to 8 eV, while Hund's rule coupling ($J_\textrm{H}$) parameter was selected to be 1 eV~\cite{ushakov2009electronic}. Integration over the Brillouin zone(BZ)  has been performed on a Mokhorst-Pack mesh $3 \times 3 \times 3$~\cite{monkhorst1976monkhorst} for primitive cell during structure optimization and over $\Gamma$-point only for bigger supercell during calculations of lattice vibrations. Equilibrium geometry has been obtained after the several stages of full structure relaxation, that include optimization of atomic position, cell shape and cell volume. The phonon dispersion and density of states (DOS) were obtained with MedeA-PHONON module, which implements a direct approach of harmonic approximation~\cite{parlinski1997first}. The so called direct approach to lattice dynamics is based on the \emph{ab initio} evaluation of forces on all atoms produced by a set of finite displacements of a few atoms within an otherwise perfect crystal. The IR and Raman spectra were calculated from the phonon frequencies at $\Gamma$-point. For the IR and Raman intensities the dielectric tensor, the Born effective charges and the Raman tensor were calculated by means of linear response calculations, also implemented in VASP~\cite{kresse1996efficient,kresse1999ultrasoft}.

\section{Results and discussions}
\subsection{Structural properties}

The CuTe$_2$O$_5$ exhibits a monoclinic structure with space group P2$_1$/c and experimental lattice parameters $a=6.871$\,\AA, $b=9.322$\,\AA, $c=7.602$\,\AA~and $\beta=109.08 ^\circ$~\cite{Hanke1973}. The lattice parameters obtained after structure relaxation were $a=6.756$\,\AA, $b=9.302$\,\AA, $c=7.353$\,\AA~and $\beta=109.08 ^\circ$. The calculated lattice parameters are all underestimated by 2-3\,\%, which is a normal DFT error. Moreover the calculated values correspond to $T=0$\,K, whereas the experimental ones  could be larger because they were measured at finite temperature.

The magnetic moment on Cu ions was found to be 0.75\,$\mu_\textrm{B}$, which well corresponds to the value 0.79\,$\mu_\textrm{B}$ obtained previously~\cite{ushakov2009electronic}. In the previous studies the dominant antiferromagnetic interaction between Cu ions was demonstrated~\cite{das2008proposed}. We have considered two type of antiferromagnetic ordering of magnetic moments on Cu-Cu dimers, which realize the P2$_1$ and Pc subgroups of crystal space group P2$_1$/c, see Fig.~\ref{fig:P21-Pc}. The magnetic ordering with P2$_1$ symmetry has smaller energy of about 2\,meV per Cu atom in comparison to Pc and was selected as a ground state for the further \emph{ab initio} calculations of vibrational properties.

\subsection{Vibrational properties}
The phonon dispersion and phonon DOS, total and decomposed on the different atom types, were calculated as described above and are presented on the Fig.~\ref{fig:PhononDOS}. One can see, that at low frequencies the dominant contribution to the lattice vibrations comes from the Cu and Te atoms, whereas the vibration modes of oxygen atoms are predominantly have higher energies. 

\begin{figure*}
\includegraphics[width=0.49\linewidth]{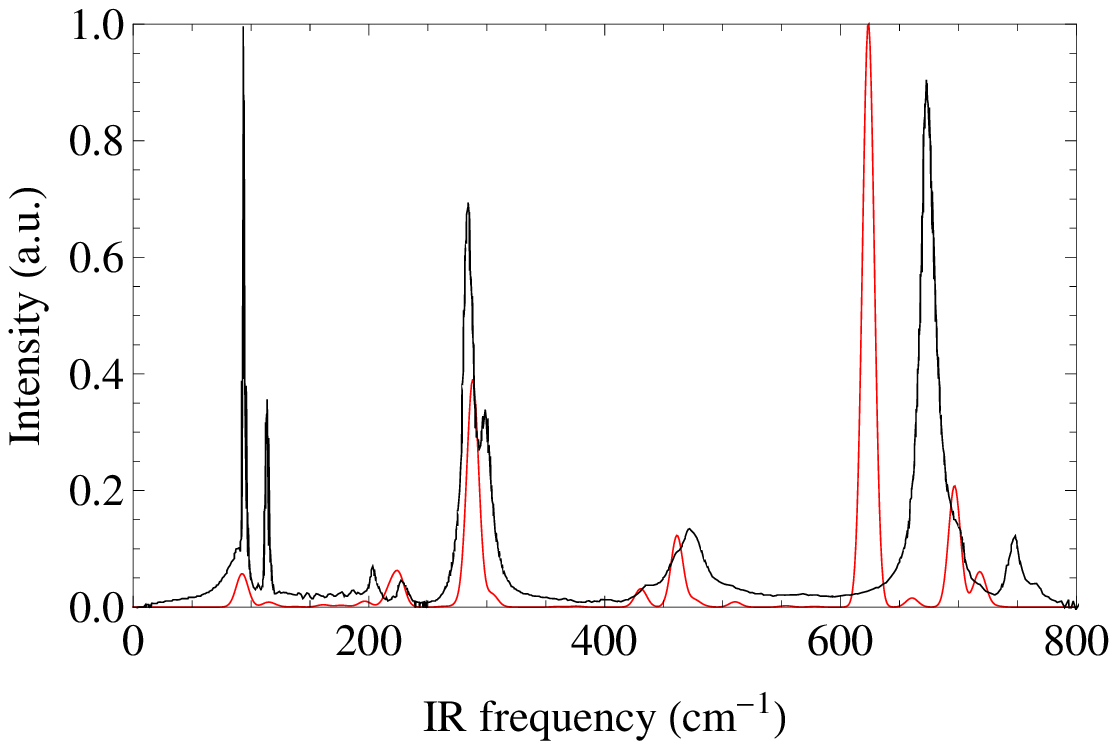}
\includegraphics[width=0.49\linewidth]{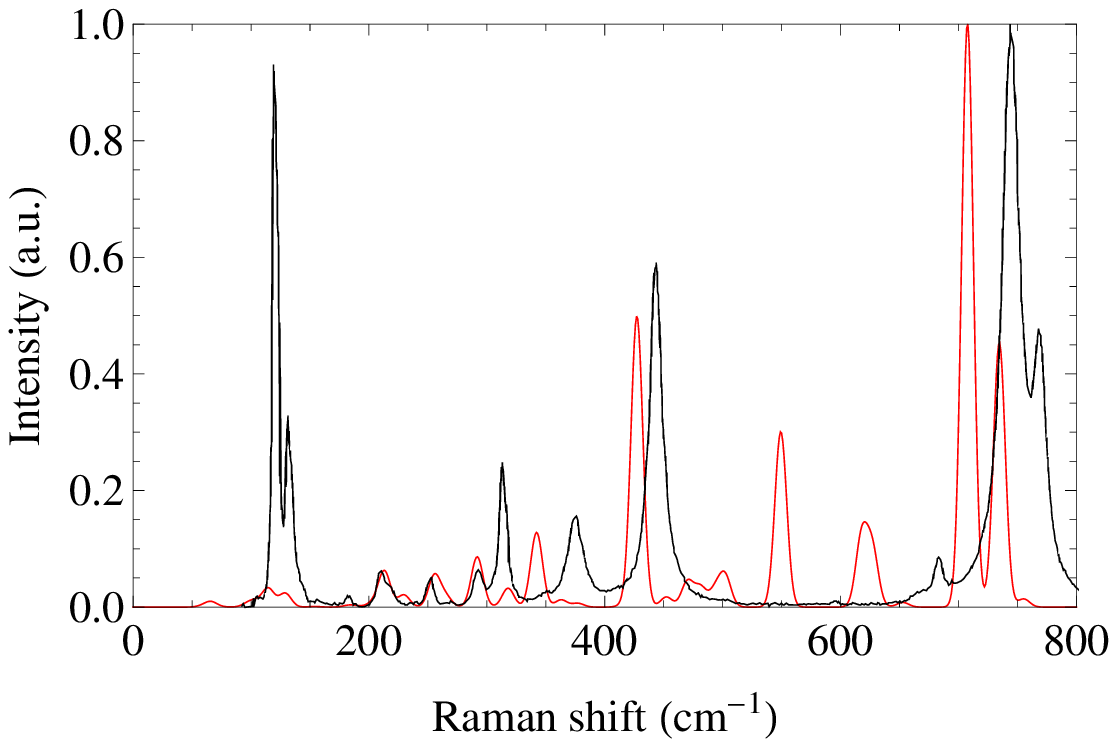}
\caption{\label{fig:IRRamanSpectras} Calculated (red) and experimental (black)~\cite{VibrationalCuTe2O5} spectra for IR and Raman measurements, normalized to the maximum peak intensity. See text for details.}
\end{figure*}

IR and Raman spectra, calculated in the present work and experimentally measured~\cite{VibrationalCuTe2O5}, are shown on the Fig.~\ref{fig:IRRamanSpectras}. The maximum peak intensities were normalized to unity while the line width of calculated spectra was selected to be 5\,cm$^{-1}$. The difference between peak intensities in experimental and calculated spectra is due to the different orientations of crystal in experiment and calculations. However, calculated positions of peaks below 400 cm$^{-1}$ on both IR and Raman spectra are well compared to the experimental ones. But there is frequencies underestimation for $\nu \ge 400$\,cm$^{-1}$ with a magnitude of about 50\,cm$^{-1}$. 
The underestimation of the frequency of the stretching modes is probably due to the overestimation of the bond length, which was reported previously for example in Ref.~\cite{ceriotti2006ab} for Te-O stretching modes above 500\,cm$^{-1}$, where the \emph{ab initio} study of the vibrational properties of crystalline TeO$_2$ was performed. 

Nevertheless, one can assume that vibrations at frequencies above 400\,cm$^{-1}$ become active only at temperatures of about 500\,K and their contribution to the heat capacity in the temperature range 10 - 100\,K is negligible.

\subsection{Heat capacity. Lattice contribution}
\label{sec:HeatCapacityPhonon}
We assume that the total heat capacity originates from
two different contributions, a lattice (phonon) contribution C$_\textrm{latt}$
due to acoustic and optical phonons and a magnetic contribution
C$_\textrm{magn}$ corresponding to the thermal population
of excited spin states.


In order to calculate the phonon contribution to the heat capacity of CuTe$_2$O$_5$ we have used approach based on the harmonic approximation, so that  heat capacity could be found as~\cite{kittel1976introduction}
\begin{equation}
C(T) = d k_\mathrm{B} \int_0^\infty g(\omega) \left(\frac{\hbar \omega}{2 k_\mathrm{B} T}\right)^2 \frac{\exp(\hbar \omega/k_\mathrm{B} T)}{\left(\exp(\hbar \omega/k_\mathrm{B} T)-1\right)^2} d\omega,
\end{equation}
where $d$ is the number of degrees of freedom in the unit cell, $g(\omega)$ is a total phonon DOS, which is shown on the Fig.~\ref{fig:PhononDOS}, $\hbar$ and $k_\mathrm{B}$ are the Planck and Boltzmann constants and $T$ is temperature.

\begin{figure}
\includegraphics[width=\linewidth]{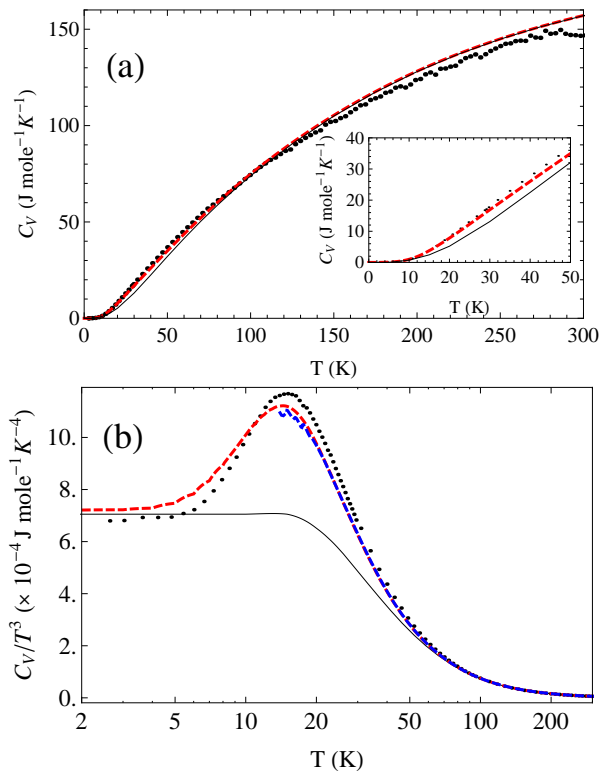}
\caption{\label{fig:Cv-exp-ph-mt} Heat capacity $C_\textrm{V}$ (a) and 
heat capacity divided by T$^3$ (b) for CuTe$_2$O$_5$ as a function of temperature: experimental data (dots), lattice
contribution (black solid line), sum of lattice and magnetic contributions (present work, red dashed line) and sum of lattice contribution with magnetic contribution, calculated with 2D-coupled dimer model~\cite{das2008proposed}.}
\end{figure}

The lattice contribution to the heat capacity calculated from the harmonic approximation is shown on the Fig.~\ref{fig:Cv-exp-ph-mt}(a) and compared to the experimentally measured total heat capacity from Ref.~\cite{Eremina2011Magnetization}. The calculated values are higher than the experimental one at high temperatures $T > 150$\,K, which could be explained by the underestimation of calculated vibration frequencies above 400\,cm$^{-1}$ that leads to the overestimation of heat capacity.

It is well known, that at low temperatures the lattice contribution to the heat capacity demonstrates the cubic dependence on the temperature. Thus
all calculated and experimental data divided by $T^3$ are plotted
as a function of $T$ in the Fig.~\ref{fig:Cv-exp-ph-mt}(b). It can
be seen, that at the temperatures below $15$\,K the calculated
lattice contribution to the heat capacity obeys this law, i.e. proportional to the $T^3$. On the
other hand, there is peak on the experimental heat capacity curve. This peak could be associated with the magnetic contribution $C_\textrm{mag}$ to the heat capacity, described in the section below.

%
%
%
%

\subsection{Magnetic contribution}
\label{sec:CvMagneticContribution}
Evidently, the magnetic contributions to the heat capacity is small compared to the
lattice contribution and a non-magnetic reference material is not
available in the case of CuTe$_2$O$_5$. Therefore, a straightforward experimental method to unambiguously extract the
magnetic contribution from the experimental data is not realize.

At the same time the magnetic contribution to the heat capacity of
the crystal containing the magnetic ions can be calculated theoretically as
\begin{equation}
C_\textrm{magn}= -T \frac {\partial^2 F}{\partial T^{2}},
\end{equation}
where $T$ is temperature, $F$ is the Helmholtz free energy
\begin{equation}
\label{eq:FreeEnergy}
F=-T\ln\sum exp(-E_n/T),
\end{equation}
where $E_n$-energy levels of the considered spin system. The
energy levels of the spin system depend on the external magnetic
field and can be described by the spin-Hamiltonian. To calculate
the magnetic contribution to the heat capacity for CuTe$_2$O$_5$
we used the following spin Hamiltonian for alternating spin chain
of ten spins
\begin{equation}
\label{eq:SpinHamiltonian}
\textrm{H}=\sum_{i \, =\, 1} ^{5} {J_1 (S_{2i-1}\cdot S_{2i})+J_2
(S_{2i}\cdot S_{2i+1})+g\beta H_{z}S_{z, i}},
\end{equation}
where $J_1$ and $J_2$ describe the isotropic exchange interaction
between spin $S_{2i}$ and its nearest neighbours $S_{2i+1}$ and
$S_{2i-1}$, the last term describes the interaction of all spins
with magnetic field $H$. 

\begin{figure}[h]
\includegraphics[width=1.1\linewidth]{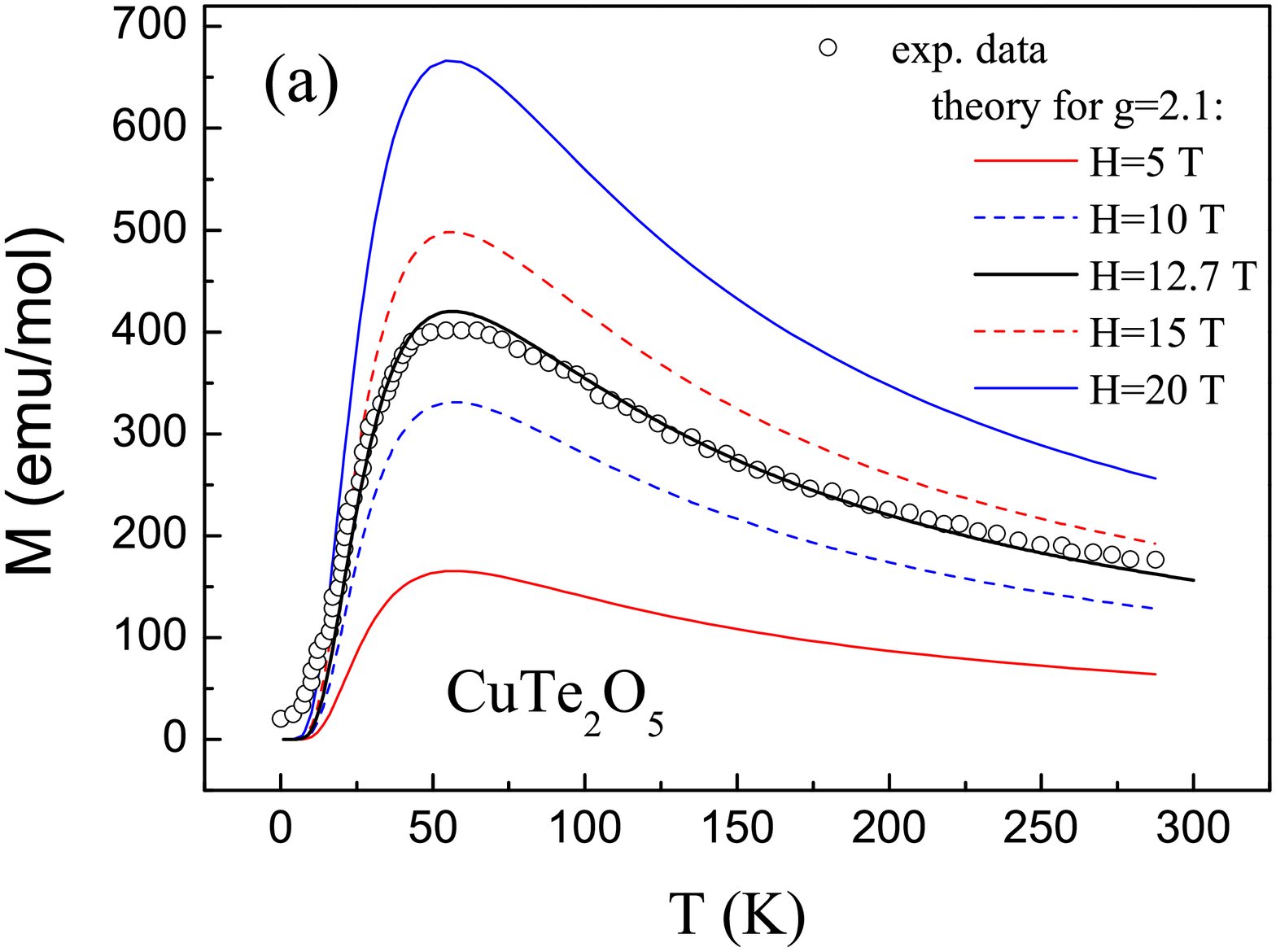}
\includegraphics[width=1.1\linewidth]{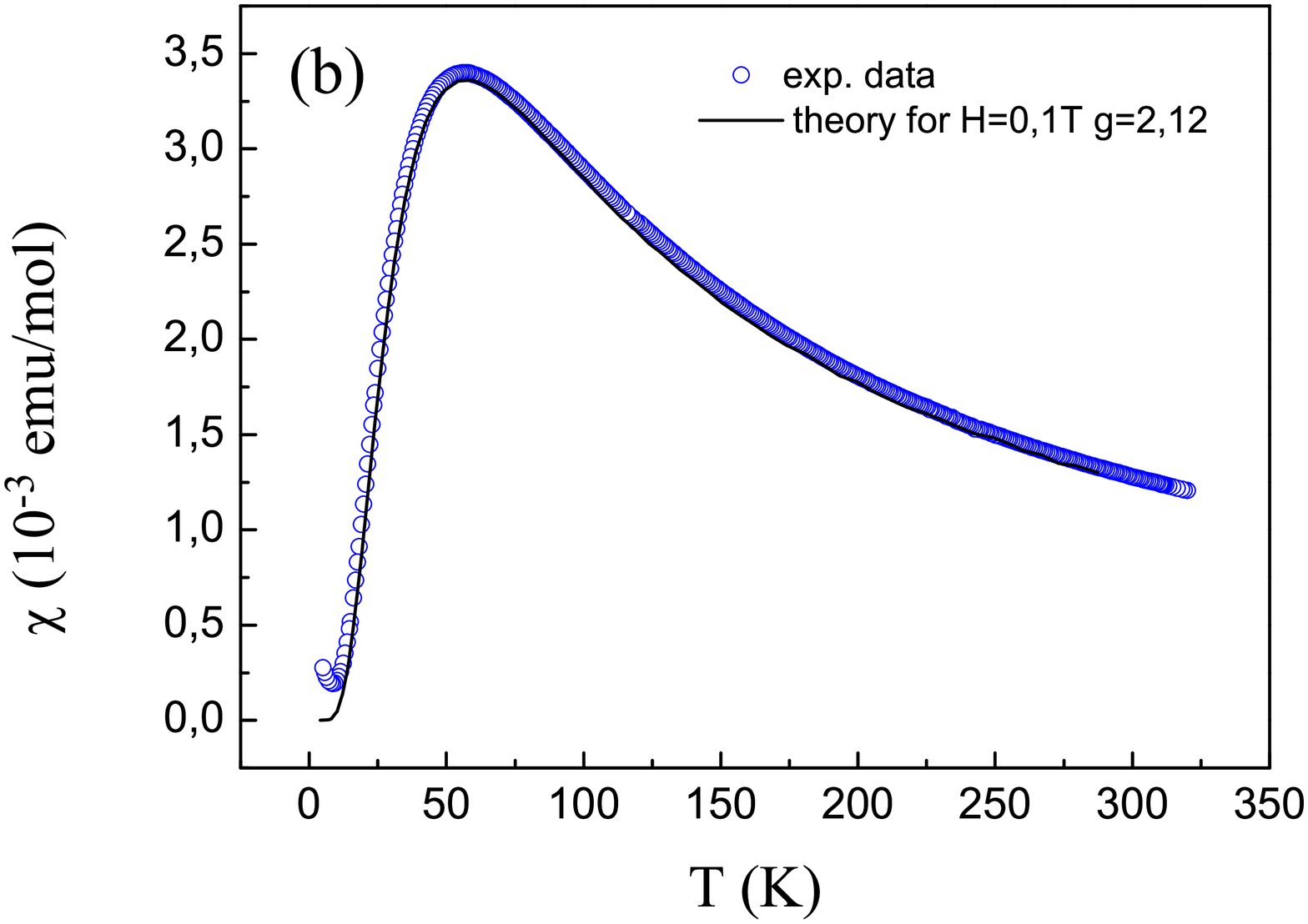}
\caption{Theoretically predicted temperature dependence of the magnetization (a) and magnetic susceptibility (b) of CuTe$_2$O$_5$
for different values of the external magnetic field $H$. Open circles correspond to the experimental data (see text).}\label{Fig:M_fields}
\end{figure}

In order to obtain these parameters, one can make a fitting of magnetization $M$ and magnetic susceptibility $\chi$ of CuTe$_2$O$_5$ to the experimental data. Using above presented Eqns.~(\ref{eq:FreeEnergy}) and~(\ref{eq:SpinHamiltonian}) one can calculate these quantities as
\begin{eqnarray}
{M} &=& - \frac {\partial F}{\partial H},\\
{\chi} &=&  \frac {\partial^2 F}{\partial H^2},
\end{eqnarray}
where $F$ is the Helmholtz free energy defined by Eq.~(\ref{eq:FreeEnergy}).
In the case of CuTe$_2$O$_5$ the values of isotropic exchange interactions are $J_1$ = 93.3\,K and $J_2$ = 40.7\,K as determined from susceptibility and ESR data \cite{deisenhofer2006structural,Eremina2008Anisotropic,Gavrilova2008Anisotropic}.
The theoretically calculated magnetization for the aforementioned values of isotropic exchange interactions  $J_1$ and $J_2$ for different values of external magnetic field is presented on the Fig.~\ref{Fig:M_fields}(a). It could be seen that magnetization $M(T)$ at $H=12.7$\,T is in good agreement
with experimentally measured data from Ref.~\cite{Eremina2011Magnetization}. The temperature dependence of the magnetic susceptibility $\chi(T)$ for $H=0.1$\,T, also coincide with experimental data from Ref.~\cite{deisenhofer2006structural} (Fig.~\ref{Fig:M_fields}(b)). So, one can conclude, that present model should describe the magnetic subsystem of CuTe$_2$O$_5$ well.

\begin{figure}
\includegraphics[width=1.1\linewidth]{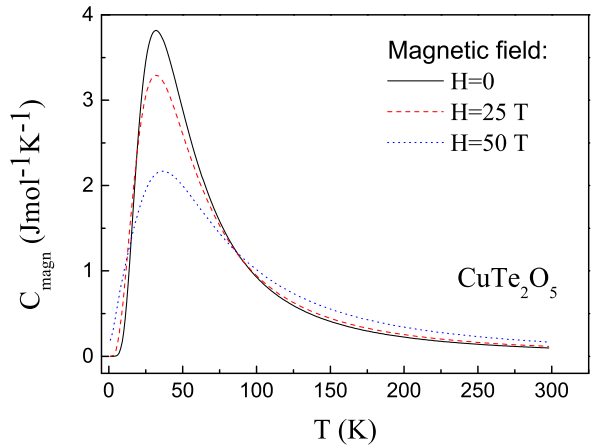}
\caption{Theoretically predicted temperature dependence of the magnetic contribution
to the  heat capacity of CuTe$_2$O$_5$ for different values of the external
magnetic field $H$.}\label{Fig:C_magn_field}
\end{figure}

As could be seen from the spin Hamiltonian (\ref{eq:SpinHamiltonian}) the energy levels $E_n$ depend
on applied magnetic field, thus the magnetic contribution to the heat capacity also depends on the values of the external magnetic
field, that is depicted on Fig.~\ref{Fig:C_magn_field}.

We added magnetic contribution to the heat capacity $C_\textrm{magn}$ to the lattice contribution $C_\textrm{latt}$, obtained in Section~\ref{sec:HeatCapacityPhonon} and compared it with experimental data~\cite{Eremina2011Magnetization}, see Fig.~\ref{fig:Cv-exp-ph-mt}. The sum of lattice contribution and magnetic contribution obtained from 2D-coupled dimer model~\cite{das2008proposed} is also shown for comparison. The agreement between theoretical and experimental results is good enough, but there is small underestimation of heat capacity by both theoretical model. 
One can suggest several probable explanation of this discrepancy: (i) additional contributions to the magnetic heat capacity; (ii) presence of impurities in the sample which affect the heat capacity measured experimentally.

\section{Conclusion}

In conclusions, the low-dimensional magnetic systems are interesting for theoretical and experimental considerations. But the theoretical calculations for such systems should be performed with a high degree of accuracy, that is achieved in the present time by using the DFT method. In our opinion, the DFT methods is suitable for calculating the physical properties of the magnetic systems, such as the lattice heat capacity, Raman and IR spectra, etc. However, to obtain the good agreement between the theoretical and experimental results, especially at low temperatures, it is necessary to consider the contribution of the magnetic subsystem, as shown for  CuTe$_2$O$_5$ in this paper. Thus, the lattice heat capacity in combination with the theoretically calculated magnetic contribution is consistent with experimental data at low temperature region, where the magnetic contribution is most important.

\section*{Acknowledgements}
The work is performed according to the Russian Government Program of Competitive Growth of Kazan Federal University.

\section*{Bibliography}
\renewcommand\refname{}

\end{document}